\documentclass[twocolumn,tightenlines,pra,aps,superscriptaddress]{revtex4}
\usepackage{graphicx}

\newcommand\foreign[1]{{\it #1\spacefactor=1000}}
\newcommand\eg{\foreign{e.g.}}
\newcommand\ie{\foreign{i.e.}}
\newcommand\ibid{\foreign{ibid}}
\newcommand\kbar{\ensuremath{\mathchar '26\mkern -9muk}}

\newcommand\Denv {{D_{\mbox{\scriptsize env }}}}
\newcommand\Dcl  {{D_{\mbox{\scriptsize cl  }}}}
\newcommand\Dinit{{D_{\mbox{\scriptsize init}}}}
\newcommand\Lim{\mathop{\rm Lim}}

\begin{document}


\preprint{LA-UR-01-1909}
\title{The $\delta$-function-kicked rotor: Momentum diffusion and the quantum-classical boundary%
       \vbox to 0pt{\vss
                    \hbox to 0pt{\hskip-75pt\rm LA-UR-01-1909\hss}
                    \vskip 25pt}
      }

\author{Tanmoy Bhattacharya}
\author{Salman Habib}
\author{Kurt Jacobs}

\affiliation{T-8, Theoretical Division, MS
B285, Los Alamos National Laboratory, Los Alamos, New Mexico 87545}

\author{Kosuke Shizume}
\affiliation{University of Library and
Information Science, 1-2 Kasuga, Tsukuba, Ibaraki 305, Japan}

\date{\today}

\begin{abstract}
We investigate the quantum-classical transition in the delta-kicked
rotor and the attainment of the classical limit in terms of
measurement-induced state-localization.  It is possible to study the
transition by fixing the environmentally induced disturbance at a
sufficiently small value, and examining the dynamics as the system is
made more macroscopic.  When the system action is relatively small,
the dynamics is quantum mechanical and when the system action is
sufficiently large there is a transition to classical behavior.  The
dynamics of the rotor in the region of transition, characterized by
the late-time momentum diffusion coefficient, can be strikingly
different from both the purely quantum and classical results.
Remarkably, the early time diffusive behavior of the quantum system,
even when different from its classical counterpart, is stabilized by
the continuous measurement process.  This shows that such measurements
can succeed in extracting essentially quantum effects.  The transition
regime studied in this paper is accessible in ongoing experiments.
\end{abstract}


\maketitle

\vskip2pc

\section{Introduction}

Explorations of the transition from quantum to classical behavior in
nonlinear dynamical systems constitute one of the frontier areas of
present theoretical research.  The recently realized possibility of
carrying out controlled experiments in this
regime~\cite{exp1,exp2,exp3} has added greatly to the impetus for
increasing our understanding of this transition, quite aside from the
undoubtedly fundamental importance of the subject.  The point at issue
is not so much the status of formal semiclassical approximations in
the sense of taking the mathematical limit $\hbar\rightarrow 0$, but a
description and understanding of the processes that take place in
actual experiments on the dynamical behavior of {\em observed} quantum
systems.  Quantum decoherence and conditioned evolution arising as a
consequence of system-environment couplings and the act of observation
provide a natural pathway to the classical limit as has been
demonstrated quantitatively in Ref.~\cite{bhj} (see also
references~\cite{clmeas1}) and is reviewed in the next section.

Comparison of the dynamics of closed quantum and classical systems in
explicit examples has shown a variety of behaviors.  At the one
extreme, there exist systems where quantum and classical averages
track each other for long times~\cite{long}, and at the other extreme,
there are systems in which the averages break away from each other
decisively at finite times~\cite{short}.  Since even this second class
of systems do attain classical dynamics when the action is
sufficiently large, they undergo a clear transition from a `quantum'
to a `classical' behavior as the parameters of the system controlling
its action are varied, and are of particular interest to a study of
the quantum-classical transition.  The quantum delta-kicked rotor
(QDKR) is a well-studied example of this class of system and it is our
purpose here to investigate the quantum-classical transition in this
system within the framework based on the theory of continuous
measurement presented in Ref.~\cite{bhj}.  As will be demonstrated
below, the quantum-classical transition regime possesses some
remarkable features which are well within the reach of present-day
experiments.

One of the remarkable features of the QDKR which distinguishes it from
its classical counterpart is the behavior of its late time momentum
diffusion constant, $D_p \equiv \Lim_{t\to\infty} d \langle p^2(t)
\rangle / dt$.  In the classical case, this quantity attains a
constant value, $\Dcl$, whereas for the QDKR, it falls to zero.  This
latter phenomenon is termed dynamical localization~\cite{dloc}.  In
addition, the QDKR also displays a nontrivial variation of the
early-time intrinsic momentum diffusion coefficient, $\Dinit$, as a
function of the stochasticity parameter~\cite{shep}, and strong
resistance to decoherence~\cite{qdkrdeco,cohen,hjmrss,milner,steck,
scalings}.  This last feature relates to the fact that it is possible
for external noise and decoherence to break dynamical localization in
the sense that the late-time quantity $D_p\neq 0$, nevertheless $D_p$
remains small and far from its classical value unless very strong
noise strength is employed.  Finally, the QDKR possesses the added
attraction that it can be studied via atomic optics experiments
utilizing laser-cooled atoms and allowing some degree of control of
the coupling to an external environment via spontaneous emission
processes~\cite{exp2}.  This system consists of noninteracting atoms
in a magneto-optical trap (MOT); on turning on a detuned standing
light wave with wave number $k_L=2\pi/\lambda$, the atoms scatter
photons resulting in an atomic momentum kick of $2\hbar k_L$ with
every such event.  Internal degrees of freedom can be eliminated since
the large detuning precludes population transfer between atomic
states.  The resulting dynamics is restricted to the momentum space of
the atoms and can be described by simple effective Hamiltonians, the
Hamiltonian for the QDKR being one of the implementable examples.

Our main consideration is a systematic analysis of the
quantum-classical transition induced via position measurement for the
non-dissipative delta-kicked rotor using the late-time momentum
diffusion rate as a diagnostic tool.  Even though quantum dynamical
localization typically suppress late time diffusion ($D_p=0$ at late
times), we find that a small measurement strength (i.e., weak noise)
can stabilize an intrinsically quantum early time effect (the
enhancement of the initial quantum diffusion rate predicted by
Shepelyansky~\cite{shepscaling}) and produce a nonzero final diffusion
rate much larger than the classical value.  The predicted effect is
large and should be observable in present experiments studying the
quantum-classical transition in the QDKR.  We also comment on the
behavior of the diffusion coefficient near a quantum resonance and on
the nontrivial nature of the approach to the classical noise-dominated
value for the diffusion rate as the noise induced by the measurement
is increased.  Our theoretical results have been confirmed recently by
a more detailed analysis relevant to specific experimental
realizations~\cite{nz}.

The plan of the paper is as follows.  In Section II, we provide a
short review of recent work on continuous measurement and the
quantum-classical transition, moving on to the specific case of the
QDKR in Section III.  In Section IV, we explain how the late-time
diffusion coefficient provides a window to investigate the
quantum-classical transition in this system and in Section V we
describe the detailed nature of the transition.  Section VI is a short
conclusion.

\section{Continuous Measurement and the Quantum to Classical
Transition} 

Macroscopic mechanical systems are observed to obey classical
mechanics.  However, the atoms which ultimately make up the
macroscopic systems certainly obey quantum mechanics.  Since classical
and quantum evolution are different, the question of how the observed
classical mechanics emerges from the underlying quantum mechanics
arises immediately.  This emergence, referred to as the quantum to
classical transition, is particularly curious in the light of the fact
that classical mechanical trajectories are governed by nonlinear
dynamics that often exhibit chaos, whereas the very concept of a phase
space trajectory for a closed quantum system is ill-defined, and any
signatures of chaos are, at best, indirect.

If quantum mechanics is really the fundamental theory, then one must
be able to predict the emergence of (the often chaotic) classical
trajectories by describing a macroscopic object with sufficient
realism, but fully quantum mechanically.  Sufficient ingredients to
perform such a description have now been found~\cite{bhj,clmeas1}.
The solution has involved the realization that {\em all} real systems
are subject to interaction with their environment.  This interaction
does at least two things.  First, it subjects the system to noise and
damping~\cite{CL,qnoise} (as a consequence all real classical systems
are subject to noise and damping - even if very small), and second,
the environment provides a means by which information about the system
can be extracted (effectively continuously if desired), providing a
measurement of the system~\cite{cm}.

Since observation of a system is essential in order that the
trajectories followed by that system can be obtained and analyzed
(this being just as true classically as quantum mechanically), it may
be expected that this process must be included in the description of
the macroscopic system in order to correctly predict the emergence of
classical trajectories.  An example of an environment that naturally
provides a measurement is that of the (quantum) electromagnetic field
with which the system is surrounded.  Monitoring this environment
consists of focusing the light which is reflected from the system,
allowing the motion to be observed.  If the environment is not being
monitored, then the evolution is simply given by averaging over all
the possible motions of the system.  (Classically this means an
average over any uncertainty in the initial conditions, and over the
noise realizations.)

It has now been established quantitatively that continuous observation
of the position of a single quantum mechanical degree of freedom, is
sufficient to correctly predict the emergence of classical motion when
the action of the system is sufficiently large compared to $\hbar$.
In particular, inequalities have been derived involving the strength
of the environmental interaction and the Hamiltonian of the system,
which, if satisfied, will result in classical motion~\cite{bhj} (see
also~\cite{clmeas1}).  These inequalities, therefore, refine the
notion of what it means for a system to be macroscopic.

A detailed explanation of the reason that continuous measurement
induces the qantum-to-classical transition may be found in
Refs.~\cite{bhj,bhj2,bhjLA}.  To recapitulate that analysis, the
emergence of classical behavior arises from simultaneous satisfaction
of two counteracting constraints.  The first is that a sufficiently
strong observation process is needed to maintain localization of the
particle in phase space, and, thereby, produce classical motion of the
centroid of the Wigner function~\cite{bhj} from Ehrenfest's theorem.
On the other hand, even for a localized distribution, a strong
measurement introduces noise into the system, and this needs to be
bounded to a microscopic scale.  These constraints are satisfied for
an ever increasing range of measurement strengths as the system
parameters become large enough to make the quantum unit of action,
{\it i.e.}, $\hbar$, negligible.  Thus, when systems are sufficiently
macroscopic, they exhibit classical motion with a negligible amount of
irreducible quantum noise, a noise that, in practice, is always
swamped by classical measurement uncertainties and tiny environmental
disturbances.

As mentioned above, these arguments can be codified into a set of
inequalities.  First, to maintain enough localization to guarantee
that, at a typical point on the trajectory, one has for the force
$F(x)$, $\langle F(x) \rangle \approx F(\langle x \rangle)$, as
required in the classical limit, the measurement strength (defined
precisely in the next section), $k$, must stop the spread of the
wavefunction at the unstable points~\cite{footnote1}, $\partial_x F >
0$:
\begin{equation}
 8\eta k 
       \gg \left\vert \frac{\partial_x^2 F}{F} \right\vert
             \sqrt{\frac{\left\vert \partial_x F \right\vert}{2 m}}\,.
\label{eq:klarge}
\end{equation}

Second, as noted already, a large measurement strength introduces
noise into the trajectory.  Demanding that averaged over a
characteristic time period of the system, the change in position and
momentum due to the noise are small compared to those induced by the
classical dynamics, it is sufficient that, at a typical point on the
trajectory, the measurement satisfy
\begin{equation}
\frac{2 \left\vert \partial_x F \right\vert}{\eta s} \ll \hbar k
   \ll \frac{\left\vert \partial_x F \right\vert s}{4}\,,
\label{eq:ksmall}
\end{equation}
where $s$ is the typical value of the action~\cite{footnote2} of the
system in units of $\hbar$ and $\eta$ ranges from zero to unity and
characterizes the efficiency of the measurement (for the measurements
considered in this paper, $\eta=1$).  Obviously as $s$ becomes large,
this relationship is satisfied for an ever larger range of $k$, and
this defines the classical limit.

This understanding of the quantum-to-classical transition in terms of
quantum measurement which has emerged within the last ten years is, of
course, completely consistent with the mechanism usually referred to
as decoherence, since averaging over the results of the measurement
process gives the same evolution as an interaction with an unobserved
environment (in particular, the environment through which the system
is being monitored) in which the environment is traced over.  Thus,
the treatment in terms of measurement is actually a microscopic
analysis of the process of decoherence.  However, examining the
measurement process allows us to obtain an understanding of why it is
that decoherence causes classical motion to emerge, and also allows us
to realize the trajectories themselves, something that is impossible
when the environment is merely traced over.  The knowledge of the
dynamics of the individual quantum trajectories provides new
information not available from a traditional decoherence analysis and,
as we show below, this more microscopic information can be helpful in
understanding phenomena even at the level of expectation values.

With this understanding of the mechanism of the emergence of classical
motion, it becomes pertinent to ask the question, how does the
dynamics of a particular system change as it is made more macroscopic?
That is, what happens to the dynamics as it passes through the
transition from quantum motion (when its action is very small), to
classical motion (when its action is sufficiently large).  In the
following sections we address this question for the delta-kicked
rotor.

\section{QDKR Under Continuous Observation}

The Hamiltonian controlling the evolution of the QDKR is
\begin{equation}
\tilde{H}(p',q',t')={1\over 2}{p'^2\over m}+\alpha_0 cos(2k_Lq')\sum_n
\delta(t'-nT).
\end{equation}
It is more convenient to study this system by rescaling variables, in
terms of which the new Hamiltonian becomes~\cite{scalings}
\begin{equation}
H(p,q,t)={1\over 2}p^2+\kappa \cos q \sum_n\delta(t-n).
\end{equation}
where $q$ and $p$ are dimensionless position and momentum, satisfying
$[q,p]=i\kbar$, for a dimensionless $\kbar=4\hbar k_L^2T/m$, and
$\kappa$ is the scaled kick strength.  Following the experimental
situation, we will use a typical value of $\kappa=10$ and open
boundary conditions on $q$.  The value of $\kbar$ is a measure of
the system action relative to $\hbar$.  When $\kbar$ is small, the
system action is large compared to $\hbar$ and the system can be
considered to be effectively macroscopic.  Conversely, when $\kbar$
is large, the system is microscopic and will behave quantum
mechanically under weak environmental interaction.  The ability to
change the system action relative to $\hbar$ (i.e., changing
$\kbar$) allows a systematic experimental study of the
quantum-classical transition.  The parameter ranges we have studied
numerically below have been chosen to be more or less typical of those
utilized in present experiments.  Finally, we note that the scaling
required to bring the equation to this dimensionless form hides the
fact that increasing the dimensionless effective Planck's constant
$\kbar$ at fixed $\kappa$ involves increasing the period between the
kicks and decreasing their strength --- thus, all else being equal,
this system is expected to behave more clasically under observation
when the kicks are harder and spaced closer in time.

The effect of random momentum kicks due to spontaneous emission can be
modeled approximately by a weak coupling to a thermal
bath~\cite{spontem}.  In current experiments the atom interacts with a
standing wave of laser light leading to a (sinusoidal) spatial
modulation of the bath coupling.  This modulation in turn produces a
corresponding spatial variation in the diffusion coefficient of the
Master equation describing the evolution of the reduced density matrix
for the position of the atom.  In the language of continuous
measurements, spontaneous emission can be regarded as a measurement of
a function of the position $x$, namely $\cos(x)$.  While one can
certainly study this class of measurement processes, in this paper we
will study the case of continuous measurements of position.  In
general, as far as the study of the quantum-classical transition is
concerned, the exact nature of the measurement process is not expected
to be important provided that it yields sufficient information to
enable the observer to localize the system in phase space.

A continuous measurement of position is described by the (nonlinear)
stochastic Schr\"odinger equation~\cite{sse}
\begin{eqnarray}
  |\tilde{\psi}(t + dt)\rangle &=& \left[ 1 - {1\over \kbar}\left(iH +
\kbar k q^2 \right) dt\right.\nonumber\\
&&\left. + 4kqR(t)dt \right] |\tilde{\psi}(t)\rangle , 
\label{nsse}
\end{eqnarray}
where the continuous measurement record obtained by the observer is
\begin{eqnarray}
R(t)dt=\langle q(t) \rangle dt + {dW/\sqrt{8k}}, 
\end{eqnarray}
$dW$ being Wiener noise.  The noise represents the inherent randomness
in the outcomes of measurements.  Aside from the unitary evolution,
this equation describes changes in the system wave function as a
result of measurements made by the observer.  The parameter $k$
characterizes the rate at which information is extracted from the
system~\cite{djj}.

Averaging over all possible results of measurements leads to a Master
equation describing the evolution of the reduced density matrix for
the system.  This Master equation has the form
\begin{equation}
  \dot{\rho} = -{i \over \kbar}[H,\rho] - k[q,[q,\rho]] 
\label{me}
\end{equation}
where the diffusion coefficient is given by $\Denv = k \kbar^2$,
with $\Denv$ the diffusion constant describing the rate at which the
momentum of a free particle would diffuse due to a thermal
environment.  From the relationship between $k$ and $\Denv$ given
above, we see that $\kbar$ determines the relationship between the
information provided about $q$, and the resulting momentum
disturbance.  The important point to note for the following is that
for a fixed $\Denv$, the measurement strength is reduced as $\kbar$
increases (conversely, for fixed measurement strength $\Denv$
increases with $\kbar$).  The stochastic Schr\"odinger equation
(\ref{nsse}) is said to represent an unraveling of the Master equation
(\ref{me}) with averages over the Schr\"odinger trajectories
reproducing the expectation values computed using the reduced density
matrix $\rho$.

All dynamical systems that one can build in the laboratory are
necessarily observed in order to investigate their motion.  For
sufficiently small $\kbar$, and a reasonable value of $k$, the
measurement maintains localization of the wavefunction, while
generating an insignificant $\Denv$.  For sufficiently localized
wavefunctions, expectation values of products of operators are very
close to the products of the expectation values of the individual
operators.  Ehrenfest's theorem then implies that the classical
equations of motion are satisfied.  On the other hand, for
sufficiently large $\kbar$ (and the same insignificant $\Denv$), $k$
is sufficiently small that the quantum dynamics is essentially
preserved.  It is the detailed nature of this transition that we now
wish to investigate.

\section{Quantum-Classical Transition and the Late-Time Diffusion
Coefficient} 

Our strategy in examining the quantum to classical transition is to
fix the level of noise (\ie, the diffusion coefficient $\Denv$)
resulting from the measurement at some sufficiently small value, so
that classical behavior is obtained for small $\kbar$ and then to
study systematically how quantum behavior emerges as $\kbar$ is
increased.  The key diagnostic is the behavior in time of the
expectation value of momentum squared, $\langle p^2(t) \rangle$.
Previous studies of the QDKR have shown that in the presence of
decoherence/measurement, dynamical localization is lost in the sense
that there exists a non-zero late-time momentum diffusion coefficient,
but that this diffusion coefficient is not necessarily the same as the
intrinsic classical diffusion coefficient $\Dcl$
\cite{qdkrdeco,cohen}.  The existence of this late-time diffusion
coefficient provides a particularly convenient means of characterizing
and studying the quantum-classical transition: The late-time diffusion
coefficient is an unambiguous, theoretically well-defined quantity
and, moreover, is also measurable in present experiments.

In the classical regime (when $\kbar$ is sufficiently small), $D_p$
attains the classical value ($D_p\simeq \Dcl$) with $\Denv \ll \Dcl$.
This should not be confused with the noisy classical limit which
arises under strong driving by the noise (large $\Denv$) in which case
$D_p \simeq \Dcl + \Denv$ ~\cite{rrw}.  At sufficiently large values
of $\kbar$, on the other hand, one expects quantum effects to be
dominant and therefore one should very nearly obtain dynamical
localization ($D_p\simeq 0$).  Thus, one of the key questions is the
behavior of $D_p$ as a function of $\kbar$ in the transition regime
in between $D_p\simeq \Dcl$ and $D_p\simeq 0$, and the variation of
$D_p$ as a function of decoherence or measurement strength as set by
the value of $\Denv$.

Analytical investigation of the transition regime is made difficult by
the nonlinearity of the dynamical equations and the lack of a small
parameter in which to carry out a perturbative analysis~\cite{fnote}.
The nonlinearity results from the fact that the measurement record
$R(t)$ drives the evolution of the wave function in equation
(\ref{nsse}) and, at the same time, is itself dependent on the
expectation value of the position.  We solved the nonlinear
Schr\"{o}dinger equations~(\ref{nsse}) numerically.

Numerical investigation of the dynamics of the environmentally coupled
QDKR requires the solution of a stochastic, nonlinear partial
differential equation.  In order to obtain the desired ensemble
averages, one needs to average over many noise realizations for each
set of parameters considered.  We implemented a split-operator
spectral algorithm on a parallel supercomputer to solve
Eq.~(\ref{nsse}), and then averaged over the resulting trajectories to
obtain the solution to Eq.~(\ref{me}). (Direct solution of
Eq.~(\ref{me}) to the desired accuracy is still a major challenge for
supercomputers.) Grid sizes for computing the wave function ranged
from $1-64K$ depending on the value of $D_{env}$ and $\kbar$.  One
thousand realizations were averaged over for each data point.  Our
numerical results for the transition regime contain a variety of
interesting phenomena which we discuss below.

\begin{figure}
\centerline{\includegraphics[width=3.25in]{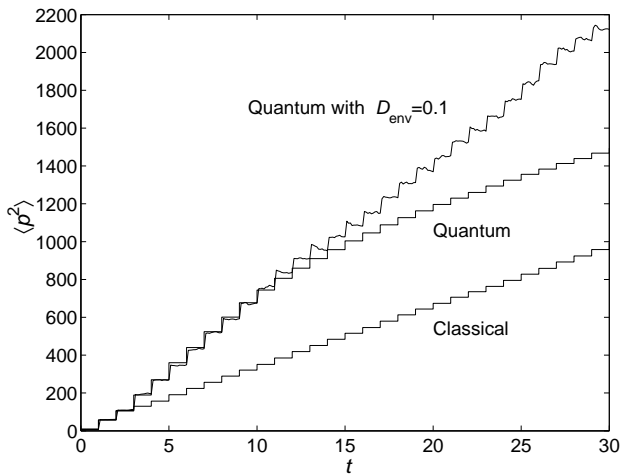}}
\caption{The spread in momentum, measured by $\langle p^2\rangle$, as
a function of time for the noiseless classical system, the noiseless
quantum system with $\kbar=3$, and the same quantum system with
$\Denv=0.1$.  To obtain the latter the Master equation has been solved
by averaging 1000 trajectories.}
\label{hbar3}
\end{figure}

\section{The Structure of the Quantum-Classical Transition}

The generic behavior of quantum trajectories is the following.  If we
construct a trajectory starting with a minimum uncertainty wavepacket,
the nonlinearity of the system dynamics tends to spread it out.
However, the localizing influence of measurement limits the spread (in
both $q$ and $p$), and a steady state is eventually attained.  The
stronger the measurement, the sooner the spread is checked.  This
behavior of individual trajectories appears to be the key to
understanding $D_p$ even though, in this case, we are interested not
in single trajectories, but in the behavior of an ensemble of these
trajectories.

In Fig.~\ref{hbar3} we plot $\langle p^2\rangle (t)$ for the classical
system, and for the quantum system with and without measurement for
$\kbar=3$.  (Using the results of~\cite{bhj}, classical behavior is
not expected unless $\Denv \ll 3$ and $\kbar \ll 2 \sqrt{\Denv}$.)  At
early times, the quantum value of the diffusion rate is much higher
than the classical value, although, consistent with dynamical
localization, this decreases with time (eventually falling to zero).
On the other hand, when the system is under observation, the initial
evolution of the system is hardly affected; it is only that the
diffusion rate reaches a constant value at about $t=10$ (for this
value of $\Denv$) at which point a purely diffusive evolution takes
over.  Thus, the measurement appears to induce a `premature'
(time-dependent) steady-state, just as it induces a `premature'
steady-state width~\cite{bhj}. Thus the diffusion rate gets frozen in
to its early time value which, in this case, is substantially larger
than the classical result.


\begin{figure}
\centerline{\includegraphics[width=3.25in]{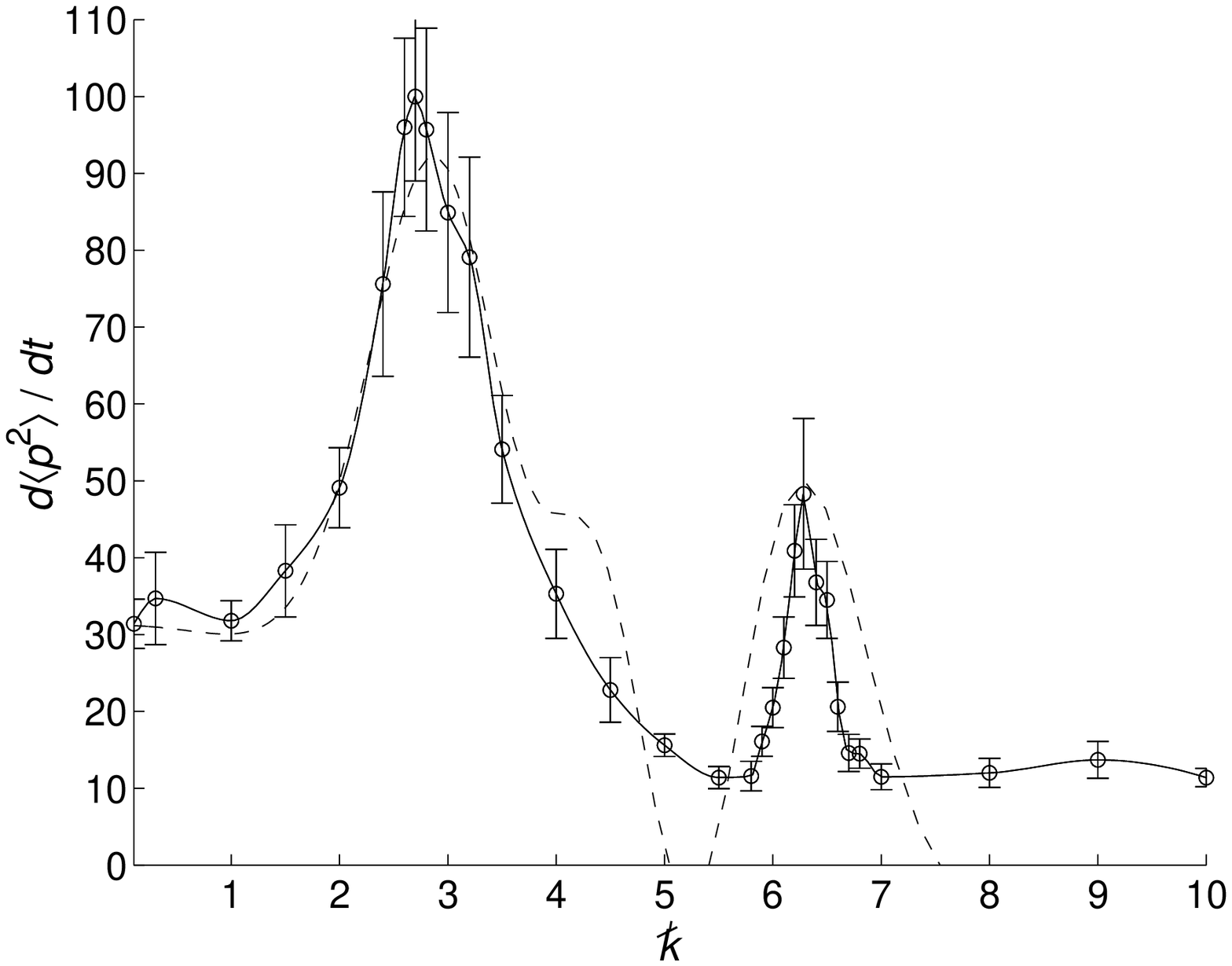}}
\centerline{\includegraphics[width=3.25in]{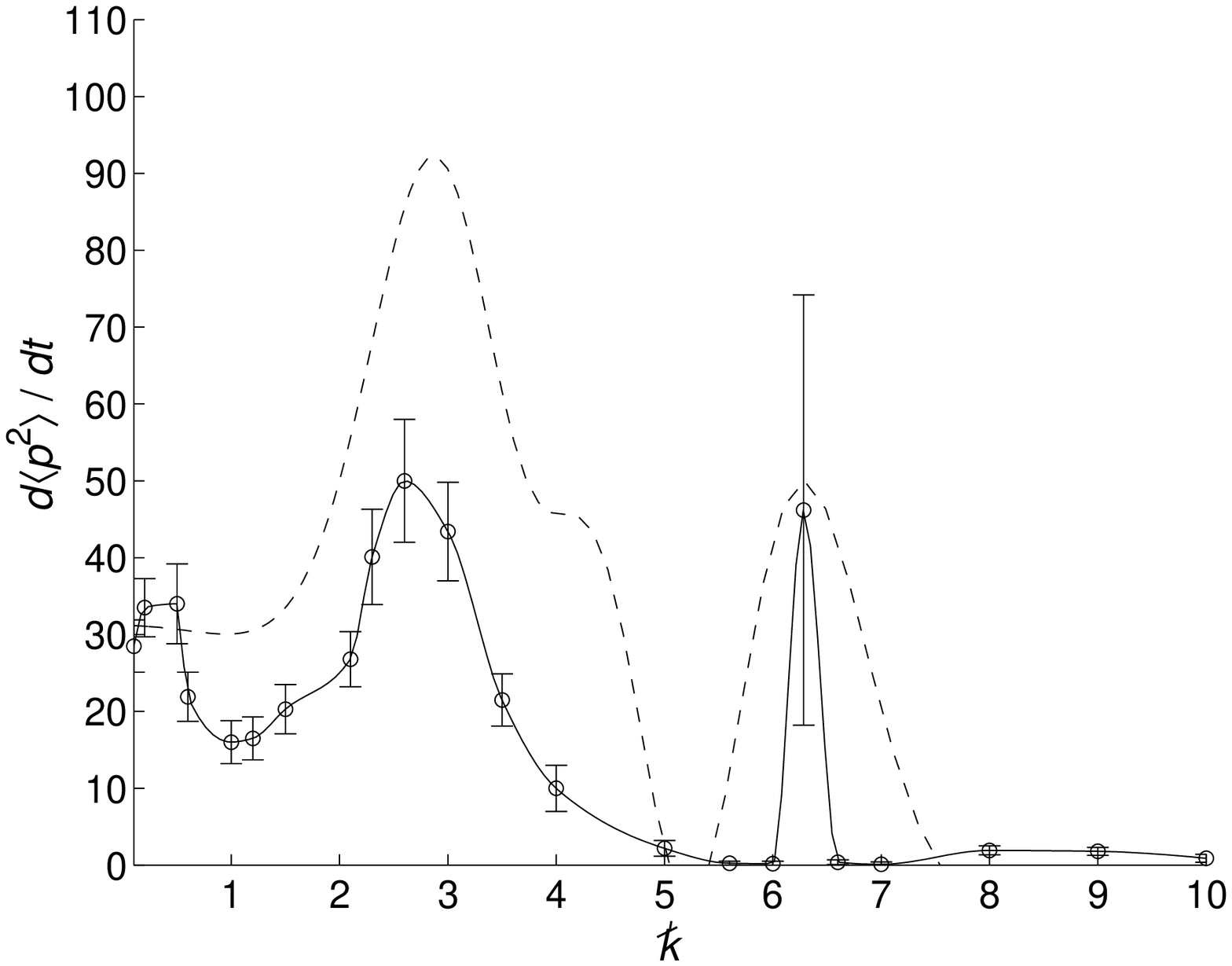}}
\caption{The late-time ($t$=30-50) diffusion coefficient as a function
of $\kbar$ for two values of the measurement induced diffusion
coefficient, $\Denv$.  The circles are the calculated points; the
solid curves are meant simply to aid the eye.  (a) $\Denv=0.1$; (b)
$\Denv=10^{-4}$.  The dashed line represents the Shepelyansky curve
discussed in the text.}
\label{tcurve}
\end{figure}

In Fig.~\ref{tcurve} we plot $D_p$ as a function of $\kbar$ for
$\Denv=0.1$ and $\Denv=10^{-4}$, with $\kappa=10$.  For small
$\kbar$, $D_p$ is essentially given by the classical value
($\Dcl=31.2$), and for large $\kbar$, $D_p$ is close to the quantum
value ($D_p=0$) when $\Denv$ is sufficiently small.  Thus we see the
expected transition from classical to quantum behavior.  However, the
transition region is surprisingly complex: the value of $D_p$ in this
region varies widely as a function of $\kbar$ and {\em rises} to more
than twice $\Dcl$ at its peak!  Another remarkable fact is that
features and placement of the transition region (as a function of
$\kbar$) is relative insensitive to the value of $\Denv$ as it
changes over three orders in magnitude.  In what follows, for
convenience we will refer to the plots in Fig.~\ref{tcurve} simply as
transition curves.

Some understanding of this complex structure can be obtained by
comparison with the early time diffusion rate for the unmeasured
quantum system as derived analytically by Shepelyansky,
\begin{equation}
\Dinit = {\kappa^2\over 2}(1 +
2J_2(\kappa_{\mbox{\scriptsize eff}}) + 2J_2^2(\kappa_{\mbox{\scriptsize
eff}}) + \cdots),   
\label{shep}
\end{equation}
where $\kappa_{\mbox{\scriptsize eff}} = 2 \kappa \sin(\kbar/2) /
\kbar$.  This approximate expression for $\Dinit$ is also plotted in
Fig.~\ref{tcurve}.  As this formula is only valid for $\kappa \gg
\kbar$, a condition not met over most of this range, we use it only as
a qualitative indicator (it is a particularly poor indicator of the
actual behavior near the quantum resonance at $\kbar=2\pi$).
Nevertheless, the trend in the data is obvious: we see that for
sufficiently large $\Denv$, the transition curve follows the early
time quantum diffusion rate fairly closely over the region of the
first peak: measurement is effective at `freezing in' the early time
value, and it is from this that the complex structure originates.

The structures in the transition region can therefore be qualitatively
understood in terms of this expression.  Comparison of Eq.~\ref{shep}
with an expression~\cite{shepscaling} for $\Dcl(\kappa)$ shows that this
initial diffusion rate in units of the square of the kick strength,
$\kappa^2$, are identical except for a renormalization of $\kappa$ to
$\kappa_{\mbox{\scriptsize eff}}$.  The classical system, however,
possesses regimes of increased diffusion due to the presence of
accelerator modes at certain values of $\kappa$.  In the quantum
system, we can scan through these values of $\kappa_{\mbox{\scriptsize
eff}}$ by tuning $\kbar$, provided $\kappa$ is large enough.  This is
the origin of the first peak in the transition region for our value of
$\kappa = 10$: at $\kbar \approx 3$, $\kappa_{\mbox{\scriptsize eff}}
\approx 6.907$, which is the position of the first such accelerator
mode.

The scaled classical diffusion constant, $\Dcl/\kappa^2$, also
increases as $\kappa \to 0$.  This leads to an interesting behaviour
in the quantum expression because by choosing $\kbar = 2 \pi$, we can
tune $\kappa_{\mbox{\scriptsize eff}} = 0$ even when $\kappa$ remains
non-zero.  Correspondingly the quantum system with this value of
$\kbar$ shows an enhanced early diffusion which has no simple
classical counterpart.  In fact, in a purely quantum mechanical
approach, this very narrow quantum resonance~\cite{steck} arises due
to quantum mechanical interference effects with no classical analog.
It is remarkable that these inherently quantum mechanical effects
survive and, in fact, are stabilized by continuous measurement and the
associated decoherence in the Master equation~(\ref{me}).  This
counter-intuitive behavior results in part from the fact that the QDKR
strongly resists decoherence to classicality as explained in
Ref.~\cite{hjmrss}.

For smaller values of $\Denv$, the transition curve drops below the
Shepelyansky predictions for the diffusion rate, which is consistent
with the notion that the weaker measurement takes longer to stabilize
the falling quantum diffusion rate.  As is evident from
Fig.~(\ref{hbar3}) locking in at a later time on the noiseless quantum
curve will clearly produce a smaller value of the diffusion
coefficient.

\begin{figure}
\centerline{\includegraphics[width=3.2in]{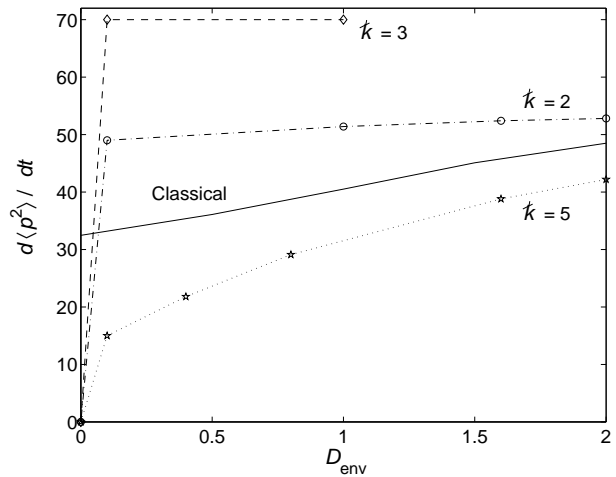}}
\caption{The late time momentum diffusion coefficient $D_p$ as a
function of the diffusion coefficient $\Denv$ in the Master equation
for the QDKR.  Results at different values of $\kbar$ show the
nontrivial nature of the approach to the classical noise-dominated
result (solid line).}
\label{DpDenv}
\end{figure}

One consequence of this complex behavior is that, in the transition
region, the cross-over from the classical to quantum regimes can also
lead to effects that are quite counter-intuitive.  An example of such
behavior is exposed by plotting the late-time $D_p$ as a function of
$\Denv$ for different values of $\kbar$ as we have done in
Fig.~\ref{DpDenv}.  One interesting open question that can be
addressed this way is whether the quantum evolution, as a function of
$\Denv$, first goes over to the classical limit (with small noise) or
reaches the classical value only in the fully noise dominated limit.
At sufficiently small $\kbar$ and a finite $\Denv$, it follows from
the results of Ref.~\cite{bhj} that the classical limit will exist and
thus the quantum evolution will go over to the small-noise classical
limit.  However, as Fig.~\ref{DpDenv} shows, in the intermediate
regime this is not the case.  Indeed, for a range of values of
$\kbar$, an inversion of what is usually expected occurs: The system
diffuses slower (intuitively, a `more quantum' behavior) at {\em
smaller} values of $\kbar$ (\eg, $\kbar=3$ vs.\ $\kbar=5$ in
Fig.~\ref{DpDenv}).

In summary, we would like to emphasize certain important points
regarding the transition curve.  The first is that examining only the
transition curve, one might conclude that the classical limit has been
achieved when the classical value of $D_p$ has been reached (\eg\ at
$\kbar\approx 0.2$ for $\Denv=10^{-4}$), especially since it remains
at this value for smaller $\kbar$.  However, examining the position
probability distribution of the particle for a typical trajectory with
$\kbar=0.2$, and $\Denv=10^{-4}$, we find that far from being well
localized, the particle position is spread significantly over four
periods of the potential, and the distribution contains a great deal
of complex structure.  As a result, the individual trajectories, which
are in principle measurable, are still far from
classical~\cite{fnote2}.  Evolving for small $\kbar$ reveals that
true classical motion emerges at the trajectory level for
$\Denv=10^{-4}$ only when $\kbar\lesssim 10^{-3}$, as expected from
the conditions in Ref.~\cite{bhj}.  The second point is that since the
transition curves rise above the classical value in the intermediate
regime, they necessarily cross this value again during their descent
into the quantum region.  Hence it is important not to sample the
curve only in a limited region, in which case one could mistakenly
conclude that the transition from quantum to classical behavior of the
diffusion constant had already taken place.

Experimental verification of our predictions should be within reach of
the present state of the art.  Either spontaneous emission or
continuous driving with noise should be, in principle, sufficient to
observe the anticipated diffusive behavior in $\langle p^2(t) \rangle$
(see, \eg, Ref.~\cite{scalings}).  Measuring $\langle p^2(t) \rangle$
accurately can, however, still be complicated by problems with
spurious tails in the momentum distribution, nevertheless, these
problems can likely be overcome especially since the predicted effects
do not require the experiments to be run for long times (see
Ref.~\cite{darcy} for a recent measurement of the behavior near the
quantum resonances including decoherence effects).

\section{Conclusion}

To conclude, we reemphasize a few key points: We have shown that it is
possible to characterize the quantum-classical transition in the QDKR
by fixing the environmentally induced diffusion ($\Denv$) at some
sufficiently small value, and examining the late-time diffusion
coefficient as the size of the system is increased ($\kbar$
decreased).  In doing so, we have shown that the late-time behavior in
the transition region is strikingly complex and different from both
the classical and quantum behavior, and that this dynamics follows,
instead, the early time quantum diffusion rates.  Remarkably, the
temporary nature of the early-time quantum diffusion rates is in fact
stabilized by the continuous measurement allowing for their possible
measurement.  These predictions for a distinct, experimentally
accessible, `transition dynamics' provide an interesting area for
investigation in experiments currently being performed on the quantum
delta-kicked rotor.

\section{Acknowledgements}

We thank Doron Cohen, Andrew Daley, Andrew Doherty, Scott Parkins,
Daniel Steck, and Bala Sundaram for helpful and stimulating
discussions.  K.J. would like to thank Scott Parkins for hospitality
at the University of Auckland during the development of this work.
Numerical calculations were performed at the Advanced Computing
Laboratory, Los Alamos National Laboratory, and at the National Energy
Research Scientific Computing Center, Lawrence Berkeley National
Laboratory.


\begin{thebibliography}{99}
\bibitem{footnote1} Actually, if the nonlinearity is large on the quantum
scale, $\hbar \partial_x^2 F / F \gtrsim 4 \sqrt{\eta m \left\vert
\partial_x F\right\vert}$, $8\eta k$ needs to be much larger than
$(\partial_x^2 F(\langle x\rangle))^2 \hbar / 4 \sqrt\eta m F(\langle x
\rangle)^2$, irrespective of the sign of $\partial_x F$.  This does not
change the argument in the body of the paper.
\bibitem{footnote2} We are assuming that both $4 [m F^2 / (\partial_x
F)^2] \left\vert F/p \right\vert$ and $E \left\vert p/F \right\vert$
evaluated at a typical point of the trajectory are comparable to the
action of the system, and define this to be $\hbar s$.
\bibitem{fnote} Approximate analytic results for a related model have 
been derived by Cohen~\cite{cohen}.  However, these are not directly 
applicable to the case considered here.
\bibitem{fnote2} It is unknown whether or not the non-classicality of
the trajectories leaves a signature detectable in ensemble
measurements.
\end{thebibliography}
\end{document}